\documentclass[superscriptaddress,aps,prl,floats,twocolumn,twoside,floatfix]{revtex4}
\usepackage{epsfig} 
\usepackage{amsmath} 
\usepackage{amssymb}

\newcommand{\be}{\begin{equation}}
\newcommand{\ee}{\end{equation}}
\newcommand{\bea}{\begin{eqnarray}}
\newcommand{\eea}{\end{eqnarray}}
\newcommand{\Z}{\mathbb Z}

\def\Tr{{\rm Tr}\,}

\begin{document}
\phantom{.}
\vskip 1truecm
\title{Correlation Inequalities for Quantum Spin Systems with Quenched Centered Disorder}  
\author{Pierluigi Contucci}
\affiliation{Department of Mathematics, University of Bologna}
\author{Joel L. Lebowitz}
\affiliation{Department of Mathematics and Physics, Rutgers University}
\begin{abstract}
It is shown that random quantum spin systems with centered disorder
satisfy correlation inequalities previously proved \cite{CL} in the classical case. Consequences include monotone approach 
of pressure and ground state energy to the thermodynamic limit. 
Signs and bounds on the surface pressures for different boundary conditions
are also derived for finite range potentials.
\end{abstract}
\maketitle

Quantum spin system with quenched randomness are both 
important and theoretically challenging. They are widely used 
as models for metallic alloys in condensed matter physics,
see \cite{SS} for a review. They are also important in combinatorial optimization
problems especially in relation to quantum 
annealing procedures \cite{MN} and quantum error correcting codes \cite{JI}.

One approach to studying such macroscopic systems is 
via correlation inequalities. These are useful in many areas of statistical mechanics. 
They have been used to prove that the free energy, correlation functions and surface
free energy have well defined thermodynamic limits. They further show 
that some of these quantities approach their limit monotonically
at increasing volumes.
They have also proved important for computing 
bounds on critical temperatures and critical exponents by comparing 
lattices in different dimensions.

In a previous work \cite{CL} we have shown that, despite the presence of 
competing interactions, a general 
classical spin glass model with quenched centered disorder has a family 
of positive correlation functions.This positivity implies a
monotone behavior of the pressure with respect to the strength 
of the random interaction. From the monotonicity one can deduce 
sub-additivity of the free energy (which implies existence of its thermodynamic limit)
and also the sign of and bounds on the surface pressure for different boundary conditions. 
See also \cite{CS} for further applications of the inequalities.

The extension to the quantum case of classical correlation inequalities
may not be possible or require further conditions. 
Examples were found in \cite{HS} where some correlations 
violate the GKS inequality of type II for the isotropic Heisenberg 
model with ferromagnetic interactions, and in \cite{G2} where the GKS 
inequality of type I is violated for the anisotropic ferromagnetic 
Heisenberg model. In order to reestablish the validity of the 
GKS inequalities in quantum systems it is then necessary to impose 
further conditions on the interaction coefficients beyond positivity 
(see for instance \cite{G2} for conditions to prove GKS I in the
anisoptropic Heisenberg model with zero magnetic field and \cite{GG} for 
the conditions to prove GKS II).

Here we show that quantum systems with quenched centered disorder
do fulfill the same family of correlation inequalities of the classical case without
any further restriction with respect to the classical case.
The result is obtained as follows.

For each finite set of points $\Lambda$ let us consider the quantum
spin system with Hamiltonian
\be
U :=
- \sum_{X\subset\Lambda}\lambda_X J_X\Phi_X + U_0
\label{sgqp}
\ee
The operators $\Phi_X$ are self-adjoint elements of the real algebra generated by the set of 
{\it spin operators}, the Pauli matrices, $\sigma_i^{(x)}, \sigma_i^{(y)}, \sigma_i^{(z)}$, $i\in \Lambda$, on the Hilbert space 
${\cal H}_X := \otimes_{i \in X}{\cal H}_i$. $U_0$ is a non random quantum Hamiltonian acting on the Hilbert space 
${\cal H}_\Lambda$. The random interactions $J_X$ are centered and mutually independent i.e. $Av(J_X)=0$ for all $X$
and $Av(J_XJ_Y)=\Delta_X^2\delta_{X,Y}$. The $\lambda$'s are numbers which tune the 
magnitude of the random interactions.
An example is the anisotropic quantum version of the nearest-neighboor 
Edwards-Anderson model with transverse field. This is defined in terms of the Pauli matrices:
\be
\Phi_{i} = \sigma_i^{z} \; ,
\ee
\be
\Phi_{i,j} = \alpha_x\sigma_i^{x}\sigma_j^{x}+\alpha_y\sigma_i^{y}\sigma_j^{y}+\alpha_z\sigma_i^{z}\sigma_j^{z} \; ,
\ee
for $|i-j|=1$ and $\Phi_X=0$ otherwise.

Our main observation is that the pressure (Gibbs free energy up to a sign) 
for $\lambda=(\lambda_{X}, \lambda_{Y},...)$:
\be
P_\Lambda({\lambda})=Av \log \Tr \exp(-U) ,
\ee
is convex with respect to each $\lambda_{X}$. We have set the inverse 
temperature $\beta=1$ since our results do not depend on its value. We
shall also drop the subscript $\Lambda$ when it is unambiguous.

The proof of convexity is straightforward. The first derivative gives in fact
\be
\frac{\partial P}{\partial \lambda_A}= Av(J_A<\Phi_A>_U)
\ee
where
\be
< C >_U :=\frac{\Tr Ce^{-U}}{\Tr e^{-U}}.
\ee
while, for the second derivative, one has (see \cite{BS}, Chapter IV, page 357)
\be
\frac{\partial^2 P}{\partial \lambda_A^2}=Av(J^2_A[<\Phi_A, \Phi_A>_U- < \Phi_A>_U^2])
\ee
where $<\cdot,\cdot>_U$ denotes the Duhamel inner product \cite{DLS}:
\be
<C,D>_U:= \frac{\Tr \int_0^1 ds\ e^{-sU}C^* e^{-(1-s)U}D}{\Tr e^{-U}}.
\ee
By using the fact that $<1,D>=< D> $ and $<C,1>=\overline{<C>}$ 
we see that
\be
\frac{\partial^2 P}{\partial \lambda_A^2}=Av(J^2_A[<\Phi_A- < \Phi_A>_U,  \Phi_A- < \Phi_A>_U>_U])\geq 0.
\ee
This yields the following result:\\

{\it For systems described by the quantum potential} (\ref{sgqp}) {\it
the following inequality holds: for all $A\subset\Lambda$ and for $\lambda_A\ge 0$ 
\be\label{qcl}
Av(J_A<\Phi_A>_U) \ge 0 \; .
\ee
}
Proof. Since the second derivative of the pressure is non negative
\be
\frac{\partial^2 P}{\partial \lambda_A^2}\geq 0.
\ee
we deduce that the first derivative
\be
\frac{\partial P}{\partial \lambda_A}=Av(J_A<\Phi_A>_U)
\ee
is a monotone non decreasing function of $\lambda_A$ (indipendently of the values
of all the other $\lambda$'s). As a consequence
we have that for $\lambda_A\ge 0$
\be
\frac{\partial P}{\partial \lambda_A}\ge Av(J_A<\Phi_A>_U)|_{\lambda_A=0}
\ee
But for $\lambda_A=0$ the two random variables $J_A$ and $<\Phi_A>_U$ are 
independent:
\be
Av(J_A<\Phi_A>|_{\lambda_A=0})=Av(J_A)Av(<\Phi_A>|_{\lambda_A=0})=0
\ee
where the last equality comes from having chosen distributions with $Av(J_A)=0$.
It also follows that for $\lambda_A\le 0$ one has $Av(J_A<\Phi_A>_U) \le 0$.

Although the consequences we are going to derive apply only to the 
case considered in (\ref{sgqp}) where $U_0$ is the sum of one body terms 
we note that the inequality (\ref{qcl}) holds for general $U_0$. 
This include the case where $Av(<\Phi_A>_U)\le 0$, as would happen in 
the case where the $J_X$ are bounded and $U$ satisfy the conditions
necessary for GKS I to hold. A different example where one exploits symmetry and translation
invariance would be the anisotropic Heisenberg model
\be
U = - \sum_{\alpha=x,y,z} K_{\alpha}\sum_{i,j}\sigma_i^{\alpha}\sigma_j^{\alpha}
-\sum_i(h+\lambda_iJ_i)\sigma_i^{z} \; ,
\ee
with centered $J_i$ and negative field $h$. 
It would also include the case $h=0$, $\Lambda \nearrow \Z^d$, 
$d\ge 3$ with minus boundary conditions and $K_{\alpha}$ positive and large.

We now consider the case where $U_0$ is a sum of one body terms, e.g. 
$U_0=-\sum_{i}\vec{h}_i\cdot\vec{\sigma}_i$. By using the same standard 
strategies of the classical spin glass case \cite{CL}
or the standard ferromagnetic interaction \cite{BS}  one can
easily deduce from (\ref{qcl}) the super-additivity of the pressure.
For a disjoint union of two regions $\Lambda=\Lambda_1\cup\Lambda_2$ one 
obtains
\be\label{mp}
P_\Lambda \ge  P_{\Lambda_1} + P_{\Lambda_2} \; .
\ee
It follows from (\ref{mp}) that the pressure is monotonically increasing
as the volume increase and hence the existence of the thermodynamic limit
(see also \cite{CGP}). Considering for instance a system on a d-dimensional square
lattice $\Z^d$, with translation invariant distributions of the random interactions,
one has that by dividing the lattice into cubes the following result holds for
free boundary conditions:
\be
p \; = \; \lim_{\Lambda\nearrow\Z^d}\frac{P_\Lambda}{|\Lambda|} \; = \; \sup_{\Lambda}\frac{P_\Lambda}{|\Lambda|}\; ,
\ee
where the supremum is a well defined function (doesn't blow up) provided the stability condition (see \cite{CGP}):
\be\label{stbl}
\sum_{X\subset\Lambda} Av(J^2_X)||\Phi_X||^2 \; \le \; c |\Lambda| \; ,
\ee 
is verified for some positive constant $c$. 
A simple bound shows that when the interactions have a finite range 
the limit does not depend on boundary conditions.

By introducing the inverse temperature in the definition of the pressure,
for instance taking all the lambdas equal to $\beta$, we can study the properties
of the ground state energy $E_\Lambda$ by relating it to the free energy. 
Since by general thermodynamic arguments (see for instance \cite{R})
\be
\lim_{\beta\to\infty} -\frac{P_\Lambda(\beta)}{\beta} \searrow E_\Lambda \; ,
\ee
one obtains
\be
E_\Lambda \le  E_{\Lambda_1} + E_{\Lambda_2} \; ,
\ee
which implies
\be
e \; = \; \lim_{\Lambda\nearrow\Z^d}\frac{E_\Lambda}{|\Lambda|} \; = \; \inf_{\Lambda}\frac{E_\Lambda}{|\Lambda|}\; .
\ee
The physical significance of a quantum disordered model is 
related to the fact that the random free energy (using now the potential with all $\lambda_X=1$)

\be
{\Pi}_{\Lambda} = \log \Tr \exp (-U_{\Lambda})
\ee
and the random ground state energy 
\be
{\cal E}_\Lambda = \lim_{\beta\to\infty} -\frac{{\Pi}_\Lambda}{\beta}
\ee
do converge, for large volumes, to the same non random object for almost all the disorder 
realizations. Following \cite{GT} and \cite{CGN} we can achieve this stronger version of the 
existence of the thermodynamic limit by observing that the condition (\ref{stbl}) entails the 
exponential version of the law of large numbers for the free and ground state energy:
\be
Prob \left(  \left| \frac{{\Pi}_\Lambda}{|\Lambda|\beta} - \frac{P_\Lambda}{|\Lambda|\beta} \right| \ge x \right) 
\le e^{-\frac{|\Lambda|x^2}{2c}} \; ,
\ee
\be
Prob \left(  \left| \frac{{\cal E}_\Lambda}{|\Lambda|} - \frac{E_\Lambda}{|\Lambda|} \right| \ge x \right) 
\le e^{-\frac{|\Lambda|x^2}{2c}} \; .
\ee
Standard probability theory (Borel-Cantelli lemma) implies that for {\it almost all} configurations of the $J$'s: 
\be
\lim_{\Lambda\nearrow\Z^d}\frac{{\Pi}_\Lambda}{|\Lambda|} \; = \; \sup_{\Lambda}\frac{P_\Lambda}{|\Lambda|} = p\; ,
\ee
and
\be
\lim_{\Lambda\nearrow\Z^d}\frac{{\cal E}_\Lambda}{|\Lambda|} \; = \; \inf_{\Lambda}\frac{E_\Lambda}{|\Lambda|} = e\; .
\ee

From inequality (\ref{qcl}) we can also deduce, by decomposing a (d+1)-dimensional hypercube $\Lambda$
into d-dimensional hypercubes, that the pressure in dimension d, $p^{(d)}$ is a non-decreasing
function of $d$ and the ground state energy a non increasing one:
\be
p^{(d)} \le p^{(d+1)} \; ,
\ee
\be
e^{(d)} \ge e^{(d+1)} \; .
\ee

In the case of finite range interactions, i.e. $\lambda_X=0$ for $|X|\ge r$ 
(e.g. the nearest neighbor case), the inequality (\ref{qcl}) leads to an
estimate of the size and sign of the surface
pressures $T_\Lambda$ i.e. the first correction to the leading term of the pressure:
\be\label{sp}
P_\Lambda = p|\Lambda|+T_\Lambda \; .
\ee
Using the methods of \cite{CL} and \cite{CG} a straightforward computation shows that 
the $T_\Lambda$ is of surface size and, as it happen for ferromagnets and 
classical spin glasses, it does depend on boundary conditions. For instance
for free ($\Phi$) and periodic ($\Pi$) boundary conditions one may show that
there are two positive constants $c_{(\Phi)}$ and $c_{(\Pi)}$ such that
\be
\label{spf}
- c_{(\Phi)} |\partial\Lambda| \le T^{(\Phi)}_\Lambda \le 0 \; ,
\ee
and
\be
\label{spp}
T^{(\Phi)}_\Lambda \le T^{(\Pi)}_\Lambda \le c_{(\Pi)} |\partial\Lambda| \; ,
\ee
where $|\partial\Lambda|$ is the area of the surface of $\Lambda$, i.e.
the number of terms in (\ref{sgqp}) which connect sites inside $\Lambda$ to
sites outside $\Lambda$.

We have shown that a disordered quantum systems fulfills a new correlation 
inequality which entails the same consequences as the first GKS inequality
and holds in full generality without any restriction with respect to the 
classical case. It would be interesting to investigate correlation inequalities
of type II (see \cite{CU,CUV} for the classical case) as well as the validity
of similar results on the Nishimori line \cite{CMN} especially in view of
the applications of the correlation inequalities to error correcting codes \cite{NM}
and their possible extension to the quantum case.

The results we have presented in this letter can of course be extended to 
quantum Hamiltonian systems with general bounded interaction. \\

{\bf Acknowledgments.} We thank Rafael Greenblatt for interesting discussions. 
P.C. thanks also Cristian Giardina, Claudio Giberti, Sandro Graffi and Walter Wreszinski. 
Research partially supported by NSF Grant DMR0802120, by AFOSR Grant AF-FA9550- 
07 and by Strategic Research Funds of University of Bologna.

\end{document}